\documentclass[showpacs,preprintnumbers,amsmath,amssymb,nofootinbib,
superscriptaddress]{revtex4}
\def\ba{\begin{eqnarray}}
\def\ea{\end{eqnarray}}
\def\lb{\label}

\def\bi{\bibitem}

\def\d{\delta}

\def\rr{\rightarrow}

\def\p{\partial}

\newcommand{\eq}{\begin{eqnarray}}
\newcommand{\eqx}{\end{eqnarray}}
\newcommand{\bfig}{\begin{figure}}
\newcommand{\efig}{\end{figure}}
\newcommand{\fr}[2]{\frac{#1}{#2}}

\newcommand{\bit}{\begin{itemize}}
\newcommand{\eit}{\end{itemize}}

\def\la{\label}

\def\bi{\bibitem}

\def\d{\partial}
\def\th{\theta}

\def\eq#1{{Eq.~(\ref{#1})}}

\reversemarginpar

\usepackage{graphicx}
\usepackage{epsfig}
\usepackage{bm}
\usepackage{amsfonts}

\begin{document} 
\title{Asymmetric 1+1-dimensional hydrodynamics in collision}

\author{A.Bialas}
\email{bialas@th.if.uj.edu.pl}
\affiliation{H.Niewodniczanski Institute of Nuclear Physics\\
Polish Academy of Sciences, Radzikowskiego 152, Krakow,
Poland\\and\\
M.Smoluchowski Institute of Physics \\Jagellonian
University, Reymonta 4, 30-059 Krakow, Poland}

\author{R.~Peschanski}
 \email{robi.peschanski@cea.fr}
 \affiliation{Institut de Physique Th\'eorique\\  CEA-Saclay,
F-91191 Gif-sur-Yvette Cedex, France}

\begin{abstract} The possibility that particle production in high-energy
collisions is a result of two asymmetric hydrodynamic flows is
investigated, using the Khalatnikov form of the 1+1-dimensional
approximation of hydrodynamic equations. The general solution is discussed
and applied to the physically appealing ``generalized in-out cascade''
where the space-time and energy-momentum rapidities are equal at initial
temperature but boost-invariance is not imposed. It is demonstrated that
the two-bump structure of the entropy density, characteristic of the
asymmetric input, changes easily into a single broad maximum compatible
with data on particle production in symmetric processes. A
possible microscopic QCD interpretation of asymmetric hydrodynamics is
proposed.
\end{abstract} 
\maketitle

\vspace{0.3cm}

\section{Introduction}

The standard phenomenological description of heavy-ion reactions is based
on the hydrodynamic evolution of the system, starting from rather early
times. This approach allows to explain many aspects of data, e.g. the
transverse momentum spectra and elliptic flow \cite{Hydr,flor}.

The hydro calculations mostly concentrate at the central rapidity region
and therefore they often leave aside the problem of longitudinal
dynamics. The usual approach for symmetric collisions ($e.g.$ for
gold-gold collisions) is to consider a rapidity symmetric medium which,
after some pre-equilbirium stage, evolves following the hydrodynamic
equations till the freeze-out temperature. The investigated solutions of
the hydrodynamic equations are naturally symmetric since they are
obtained with initial conditions which are themselves
rapidity-symmetric.

In the present paper we discuss a different hydrodynamic approach
where, starting from asymmetric initial conditions, one considers {\it
asymmetric} solutions of the hydrodynamic equations. The symmetry of the
overall solution, in particular of the rapidity distribution of
particles, is thus restored by the superposition of two conjugate
asymmetric solutions. Such a situation has been advocated to explain (i)
the forward-backward multiplicity correlations observed in
nucleon-nucleon and nucleus-nucleus reactions \cite {bz,bz1,bzw} as well
as
(ii) the sign and magnitude of the directed flow observed in 
nucleus-nucleus collisions \cite{bw}. 
It is also  natural in some
models \cite{wound, wbz,dpm, fritjof} describing the multiplicity distributions
in p-p and gold-gold reactions. Finally, it is obviously necessary for an hydrodynamic 
description of the asymmetric reactions, e.g. deuteron-gold collision
\cite{wbc}.

For this scenario to be realized, one has to show that the hydrodynamic
evolution is compatible with the existence of asymmetric components and
with their possible combination to form a reasonable overall symmetric
distribution. It is the goal of our paper to show that this is indeed
the case. We use the simplified but informative framework of the
(1+1)-dimensional approximation of the corresponding hydrodynamic flow.
In this context we i) discuss the construction of asymmetric solutions,
ii) show that the linear superposition of asymmetric components of the
multiplicity distribution in rapidity is possible to construct in an
hydrodynamic framework, more precisely of its (1+1)-dimensional
approximation by the longitudinal entropy density as a function of
rapidity.

It was shown long time ago by Khalatnikov \cite{khal} that the
hydrodynamic equations in (1+1) dimensions, considered first in the
context of high-energy collisions by Landau \cite{Lan1,bi}, can be
linearized. This is achieved by passing from kinematic variables
(position and time) into dynamic ones (temperature and rapidity).
Recently this idea was developped in a modern context \cite{Beuf,ps} and
used in the discussion of the longitudinal evolution of the symmetric
solutions. The argument of the present paper extensively exploits these
new developments and in particular the linearity property, which allows
to consider the superposition of various solutions\footnote{A recent
review of these and other investigations on Khalatnikov eq. can be found
in \cite{flor}.}.

We shall discuss how one can obtain explicit solutions of the
hydrodynamic equations which can be in phenomenological agreement with
the two-source description in terms of the wounded constituent models
\cite{wound,wbz,wbc}. In doing so, we will more generally describe the full exact solution of the (1+1)-dimensional approximation of a perfect fluid flow. Finally, we shall speculate on the possible
field-theoretical interpretation of our solutions in terms of early
asymmetric thermalization in heavy-ion high-energy collisions.

The plan of our paper is the following: in section II we recall the
derivation of the Khalatnikov equation. It allows for a general solution
of the (1+1)-dimensional approximation of the perfect fluid flow which we
derive in section III. From that general solution we discuss the
existence of conservation laws as shown in IV, and then the definition
of the initial conditions which are formulated in V. In section VI, we
derive and discuss in detail the physically appealing family of
solutions corresponding to the generalization of the ``In-Out cascade ''
scenario \cite{bj}, where the space-time and momentum rapidities are
taken equal at initial temperature. In the final section VII, we give
our conclusions and discuss the possible microscopic QCD interpretation
of asymmetric hydrodynamics.

\bigskip

\section {The Khalatnikov Equation}
\label{Khalderivation}

It is known \cite{khal,bi} that one
can formulate the problem of (1+1) hydrodynamic evolution with a linear
equation for a suitably defined potential. In this section we briefly
recall these results, recasting the calculations in
the light-cone kinematic variables, as used in \cite{Beuf,ps}
\begin{equation}
 z^\pm= z^0\pm z^1 \equiv t\pm z= \tau e^{\pm \eta}\;
\Rightarrow \; \left(\fr {\d}{\d z^0}\pm \fr {\d}{\d z^1}\right)=2
\fr {\d}{\d z^\pm} \ (\equiv2 \d_\pm)
 \label{dz},
 \end{equation} 
where
$\tau=\sqrt{z^+z^-}\ $ is the proper time and
$\eta=\frac{1}{2}\ln({z^+}/{z^-})$ is the {\it space-time}
rapidity of the fluid.
We also introduce  the
dynamical variables: $y,$  the 
{\it energy-momentum} rapidity variable and $\theta,$ the logarithm of
the inverse temperature, defined as 
 \ba
 y= \log u^+ = - \log u^- \ ;\quad\quad \theta= \log ({T_0}/{T})\ ,
\label{uplus}
\ea
 where $u^\pm =\log (u^0\pm u^1)$ are the light-cone
components of the fluid velocity and $T_0$ some given fixed
temperature. We shall choose it to be the initial temperature of the
plasma. Since plasma cools during the evolution we have always $\th>0$.

Using thermodynamic relations, one can recombine
the two relativistic hydrodynamic equations coming from the 
conservation of energy-momentum in (1+1) dimension, 
 into two equivalent ones having a  direct physical 
interpretation  \cite{khal}.
\vspace{.2cm}

(i) The relation 
\begin{equation}
\partial_+\left(e^{-\th+ y}\right)=
\partial_-\left(e^{-\th- y}\right).\label{rel1}
\end{equation}
implies the existence of a hydrodynamic potential $\Phi(z^+,z^-)$ such that:
\begin{equation}
\partial_\mp\Phi(z^+,z^-)\equiv T_0\ 
e^{-\th\pm y}= u^\pm T,\label{partialphi}
\end{equation}
meaning that {\it the
flow derives from a kinematic potential}.

(ii)  The second independent equation 
\begin{equation}
\partial_+\left(u^+s\right)+\partial_-\left(u^-s\right)=0
\label{varentropeq}
\end{equation}
corresponds to the {\it conservation of entropy along the flow}.

In order to combine usefully eqs.(\ref{partialphi}) and (\ref{varentropeq}), one introduces
 the {\it Khalatnikov potential} \cite{khal}
\begin{equation}
\chi(\theta,y)\equiv\Phi(z^+,z^-)-z^-u^+T-z^+u^-T\ ,
\label{chi0}
\end{equation}
where $z^{\pm}$ are now considered as functions of $(\theta,y)$.
This  is called the {\it hodograph} transformation expressing the
hydrodynamic equations as a function of    the dynamical variables
($\theta,y$) $via$ the Legendre transform \eqref{chi0}. Using the standard differential relations of a  Legendre transform, the
kinematic variables are recovered from the Khalatnikov potential
by the equations
\begin{equation}
z^\pm(\theta, y)=\frac{1}{2T_0}\ e^{\theta \pm y}\ \left(\partial_\theta\chi
\pm \partial_y\chi\right) . \label{zpzm}
\end{equation}

The Khalatnikov potential has the remarkable property \cite{khal,bi,ps,Beuf} to verify a
$linear$ partial differential equation which takes the form\footnote{Note that in this relation there is a sign difference comparing to
that of \cite{Bipe,Beuf}, due to the sign-difference in the
$\theta$-definition that we use in this work.}
\begin{equation}
c^2_s\,\partial_\theta^2
\chi(\theta,y)-\left[1-c_s^2\right]\partial_\theta
\chi(\theta,y)-\partial^2_y \chi(\theta,y)=0\ .
 \label{Khalatnikov}
\end{equation}
In \eqref{Khalatnikov}, $c_s$ (denoted also $1/\sqrt g$ for
further convenience) is the speed of sound in the fluid, which
will be considered as a constant in the present study.

The Khalatnikov equation has been originally  derived \cite{khal}
for the potential \eqref{chi0} and with applications to specific problems. But its range of applicability
appears to be much wider. Indeed, the transformation of a
nonlinear problem in terms of the kinematic variables into a
linear one in terms of dynamic variables has tremendous
advantages, as we shall see further. It applies to other quantities than
the Khalatnikov potential;  It allows
to obtain new solutions by arbitrary linear combinations of known
ones. Furthermore, primitive integrals and derivatives
of solutions are also solutions, as shown\footnote{The paper ref.\cite{ps} is based on the family of (1+1) dimensional ``harmonic flow'' exact solutions found in \cite{Bipe}.} in \cite{ps}.

In order to illustrate the powerfulness  of this method, let us
observe that if  $\chi(\theta,y)$
is solution of \eqref{Khalatnikov}, then the potential $\Phi,$
defined through \eqref{partialphi}, but now expressed in terms of
the hydrodynamic variables through \eqref{zpzm}, reads
\begin{equation}
\Phi(\theta,y)\equiv\Phi\{z_+(\theta,y),z_-(\theta,y)\}=
\chi(\theta,y)+\partial_\theta \chi(\theta,y)\label{PhiChi},
\end{equation}
and thus it verifies also the Khalatnikov equation
\eqref{Khalatnikov}. As a direct consequence, the physical entropy
flow as a function of rapidity also verifies \eqref{Khalatnikov}.
Indeed, one has \cite{Beuf} (see also \cite{milekhin})
\begin{equation}
\frac{dS}{dy}(\theta,y) = 2s_0T_0^g\ 
e^{-(g-1)\th}\ \partial_\theta \Phi_{} (\theta,y)\
,\label{Entr}
\end{equation}
where  we use the thermodynamic relation $s=s_0T^g=s_0T_0^g e^{-g\th}$ 
for the overall, temperature-dependent, entropy density of a perfect
fluid, recalling that by definition $c_s^2\equiv 1/g$. Since the
 entropy distribution 
 gives approximately the predicted distribution of produced
particles,
 eq.\eqref{Entr} provides 
 a rather direct link between solutions of the  Khalatnikov equation 
and  observables. Thus  \eqref{Entr} plays a central role in 
the present study.

Before going further, it proves useful  to use appropriately rescaled variables
allowing to put the  Khalatnikov equation \eqref{Khalatnikov} in a
simple and symmetric form.  Introducing
\begin{eqnarray}
 t=\fr{g-1}2\ \th\, \;\;\;\;\;x=\fr{g-1}{2\sqrt g}\ y\ ,
\label{newvar}
\end{eqnarray}
redefining $\chi(t,x)$ as a function of these reduced variables, 
and introducing the auxiliary function
\begin{equation}
Z(t,x)= e^{-t}\chi(t,x)\ ,
\la{Z}
\end{equation}
the Khalatnikov equation (\ref{Khalatnikov}) takes the simple form
\begin{eqnarray}
(\p_t^2-\p_x^2)Z=Z
\la{Khaltx}
\end{eqnarray}

\section{The solution}
\la{solution}
Equation \eqref{Khaltx} is nothing else than the 
$1\!+\!1$-dimensional Klein-Gordon equation 
(called also the Telegraphic equation \cite{bateman}). 
The  solution, for arbitrary (Cauchy)  initial conditions 
  at $t=0$, is \cite{solu}
\ba
Z=\frac12\left\{f(x+t)+f(x-t)\right\}+\frac12\int_{x-t}^{x+t}\! ds\ h(s)\ I_0(r)
+\frac12\int_{x-t}^{x+t}\! ds\ f(s)\ \frac{\d I_0(r)}{\d t}  
\lb{gensol}
\ea
where
\ba
r=\sqrt{t^2-(s-x)^2}
\ea
and $I_0(r)$ is the modified Bessel function of the first kind.

According to Eqns. (\ref{PhiChi},\ref{Entr}) and using (\ref{Z}) 
we can express the entropy distribution as 
\ba
\frac{dS}{dy}=\frac{s_0T_0^g}{2}(g\!-\!1)e^{-t}
\left\{(g+1)Z+2g\p_t Z +(g-1)\p_t^2 Z\right\}\ .
\la{finalt}
\ea

One also has for the kinematic flow variables, using \eqref{zpzm},
\ba
\tau =\frac{g\!-\!1}{2\sqrt {2 }T_0}\ e^{\fr{g\!+\!1}{g\!-\!1}\ t}
\sqrt{\left(Z+\p_tZ\right)^2-c_s^2(\p_xZ)^2};\;\;\;
\eta=y+\frac12\log\left\{\frac{Z+\p_tZ+c_s\p_xZ}
{Z+\p_tZ-c_s\p_xZ}\right\}\ .
\lb{tt}
\ea


\section {Conservation laws}
\la{Conservation}

To discuss the conservation laws we observe that it is possible to
rewrite the general solution (\ref{gensol}) in the form of an inverse
Laplace transform, implying the identity \cite{abr}
\ba
\Theta[t\!-\!|s\!-\!x|]I_0\left[\sqrt{t^2\!-\!(s\!-\!x)^2}\ \right]=
\int_{c-i\infty}^{c+i\infty}\frac{d\gamma}{2\pi i}\frac1{\sqrt{\gamma^2-1}}
\ e^{\gamma t-|s-x|\sqrt{\gamma^2-1}} \lb {foll}
\ea
with $\Re e\ c >1$.
Inserting this expression into (\ref{gensol}),  reversing the order of
integrations and using the derivative relation
\ba
\frac{\p}{\p t}\left\{\Theta[t\!-\!|s\!-\!x|]I_0\left[\sqrt{t^2\!-\!(s\!-\!x)^2}\right]
\right\}=\Theta[t\!-\!|s\!-\!x|]\frac{\p}{\p t}I_0\left[\sqrt{t^2\!-\!(s\!-\!x)^2}\right]
+\delta[t\!-\!|s\!-\!x|]  \lb{der}
\ea
one sees  that the solution (\ref{gensol}) can be written in the form
\ba
Z=\frac12\int_{c-i\infty}^{c+i\infty}\frac{d\gamma}{2\pi i}
\frac{e^{-\gamma t}}
{\sqrt{\gamma^2-1}}\int_{-\infty}^{+\infty} ds\ [h(s)+\gamma f(s)]\ 
e^{-|s-x|\sqrt{\gamma^2-1}}.  \lb{foll2}
\ea
It is not too difficult to show that (\ref{foll}) and (\ref{der}) do
indeed satisfy (\ref{Khaltx}).

Starting from (\ref{foll2}) one can derive a sum rule:
\ba
\int_{-\infty}^{+\infty}\! dx Z(t,x)=e^{-t} 
\int_{-\infty}^{+\infty}\! dx\ \frac{h(x)+f(x)}2 -e^t
\int_{-\infty}^{+\infty}\! dx\ \frac{h(x)-f(x)}2  \lb{sumrule}
\ea
where the integral
\ba
\int_{-\infty}^{\infty}\! dx\ Z(t,x)=
\int_{c-i\infty}^{c+i\infty}\frac{d\gamma}{2\pi i}
\frac{e^{\gamma t}}{\gamma^2-1}
\int_{-\infty}^{+\infty}\! dx [h(x)+\gamma f(x)]
\ea
was evaluated using the residue theorem.

From (\ref{sumrule}) one can obtain two conservation laws:
\ba
e^{-t}\ \int_{-\infty}^{+\infty}\! dx \left[Z(t,x)+\p_t Z(t,x)\right]=
 \int_{-\infty}^{+\infty}\! dx\ \frac{h(x)+f(x)}2; \lb{s}
\ea
\ba
e^{t}\ \int_{-\infty}^{+\infty}\! dx\ \left[Z(t,x)-\p_t Z(t,x)\right]= 
\int_{-\infty}^{+\infty}\! dx\ \frac{f(x)-h(x)}2. \lb{ns}
\ea

Comparing (\ref{s}) with (\ref{finalt}) one can see that (\ref{s})
implies that the total entropy is independent of $t$, i.e. independent
of the ratio $T_0/T$. This means that the total entropy does not change
during the evolution of the system. Moreover, since the freeze-out
temperature may safely be considered as constant, it also means that the
total entropy is proportional to $T_0^g$, as expected from rules of
thermodynamics.

Interestingly enough, equation \eqref{ns} furnishes another independent
conservation law for the $1\!+\!1$-dimensional hydrodynamics of a perfect
fluid. Probably due to the (1+1)-dimensional origin of the Khalatnikov
equation, it would be interesting to know its physical interpretation.
We checked the validity of both conservation laws in our numerical
applications.

\section {Initial conditions}
\la{Initial}

To discuss the consequences of the Khalatnikov eq., one has to
determine, in terms of physical quantities, (i) initial conditions at
$t=0$ (i.e. the functions $f(x)$ and $h(x)$) and (ii) the final value of
$t$ at which the solution is to be evaluated. To this end we observe that
at $t=0$ we obtain from  (\ref{gensol})
\ba
Z(x)=f(x);\;\;\;\p_tZ(x)=h(x);\;\;\; \p^2_t Z= f''(x)+f(x)\ .  \lb{init}
\ea
Introducing these formulae into (\ref{finalt}) and (\ref{tt})
we have
\ba
\left.\frac{dS}{dy}\right\vert_{t=0}=gs_0T_0^g
\left\{f(x)+h(x) +\frac{g-1}{2g}f''(x)\right\}\ ,
\la{final}
\ea
and
\ba
\tau_0(x) = \fr 1{2 T_0} 
\sqrt{\left\{[f(x)+h(x)]^2-[f'(x)]^2/g\right\}};\;\;\;
\eta_0(x)=\fr{2\sqrt g}{g-1}\ x \left(= y\right) +\frac12\log\left\{\frac{f(x)+h(x)+c_sf'(x)}
{f(x)+h(x)-c_sf'(x)} \right\}\lb{tau2}
\ea
where $(\tau_0(x),\eta_0(x))= \left.(\tau,\eta)\right\vert_{t=0}$.

To determine the final value of $t$ we have to fix the physical
conditions at which the system changes into free-moving hadrons. 
It seems reasonable to assume  that this happens at a fixed freeze-out
temperature $T=T_f$. Then the final value of $t$ at which the physical
solution should be evaluated is 
\ba
t\to t_f=\log \frac{T_0}{T_f} + \log\frac{g\!-\!1}2\ .  \lb{tfx}
\ea
We thus conclude that the initial temperature $T_0$ is needed to be defined in order to 
determine the final value of $t$ at which the physical solution is to be
evaluated.

If $T_0$ is fixed then eq.(\ref{tau2}) determines the proper time
$\tau_0(x)$ at which the evolution starts and eq.(\ref{tt})
gives the proper time at which freeze-out takes place.
To illustrate these features, in the next section we discuss a simple
while physically appealing case, when $f(x)=0.$

\section{Generalized in-out cascade}
\la{Generalized}

When $f(x) \equiv 0$, we have
\ba
Z=\frac12\int_{x-t}^{x+t}\! ds\ h(s)\ I_0(\sqrt{t^2-(s-x)^2})\ .
\lb{ZZZ}
\ea
with $t=t_f$ given by (\ref{tfx}). Using  this we also derive
\ba
\p_t Z= \frac12 \{h(x\!+\!t)+h(x\!-\!t)\} +\frac12\int_{x-t}^{x+t}\! ds\ h(s)\ 
\frac{\p I_0(\sqrt{t^2-(s-x)^2})}{\p t}
\ea
and
\ba
\p_t^2 Z= \frac12 \{h'(x\!+\!t)-h'(x\!-\!t)\} + \frac t4\left\{h(x\!+\!t)+h(x\!-\!t)\right\}
+\frac12\int_{x-t}^{x+t}\! ds\ h(s)
\frac{\p^2 I_0(r)}{\p t^2}
\ea
where we have used the formula
\ba
\frac{\p I_0(r)}{\p t}=I_0'(r)\frac t r=\frac {t I_1(r)}r \rr\frac t 2 
{\rm \ when\ }r\to 0\ .
\ea

To see the physical meaning, we observe that when
(\ref{ZZZ}) is introduced into (\ref{zpzm}) one has
\ba
z^\pm(0, x)=\left\{\frac{g-1}{4T_0}\ h(x)\right\}\  u^\pm;\;\;\;\;\Rightarrow\;\;\;\;
 \eta_0=y\vert_{t=0}\ ,
 \lb{zsimu}
\ea
i.e. at $t=0$ the position $z^\pm$ is aligned along the velocity
$u^\pm$. This can be interpreted as  a generalization of the in-out cascade
\cite{bj} for which $\eta = y$ at all temperatures leading 
to boost-invariance hydrodynamics (or ``Bjorken flow''). 
In  our generalized  in-out cascade, the  relation $\eta = y$ is  
valid at the initial stage $t=0$ only. The boost invariance is violated 
(unless $h(x)$ is a constant) because the  ratio $u/z$ depends on rapidity. 

Eqs. (\ref{final}) and (\ref{tau2}) reduce to
\ba
\left.\frac{dS}{dy}\right\vert_{t=0}=\frac{s_0T_0^g}{2}\ h(x);\;\;\;
\tau_0(x)=\frac{g-1}{4T_0}\  h(x) ;\;\;\; \eta_0=y\ .
\la{final0}
\ea

To show a specific example, let us consider the following simple
 parametrization of the intitial entropy distibution at $T=T_0 (t=0):$
\ba
h(x) = (1-x/x_{max})^a\ (1+x/x_{max})^b\ \exp{[(x/x_{max}-1)/c}]\ ,
\la{hx}
\ea
where $x_{max}=t_{max}=\log T_0/T_f$ and $a,b,c$ are constants characterizing the initial entropy profile.
In the plots we keep $a=b=2$ for simplicity and vary $c = \{0.3,0.4,0.5\}.$

\bfig
{\includegraphics[height=0.25\textheight]{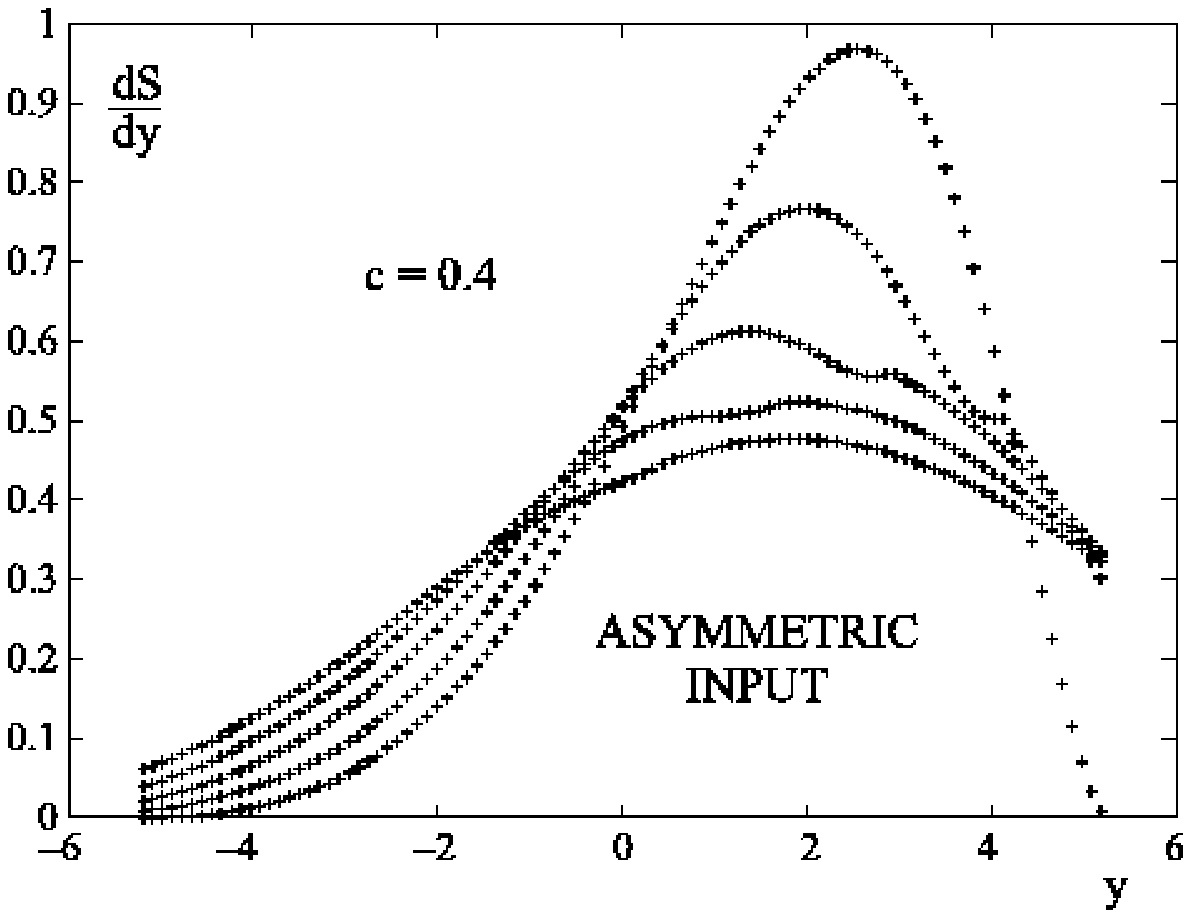}}
{\includegraphics[height=0.25\textheight]{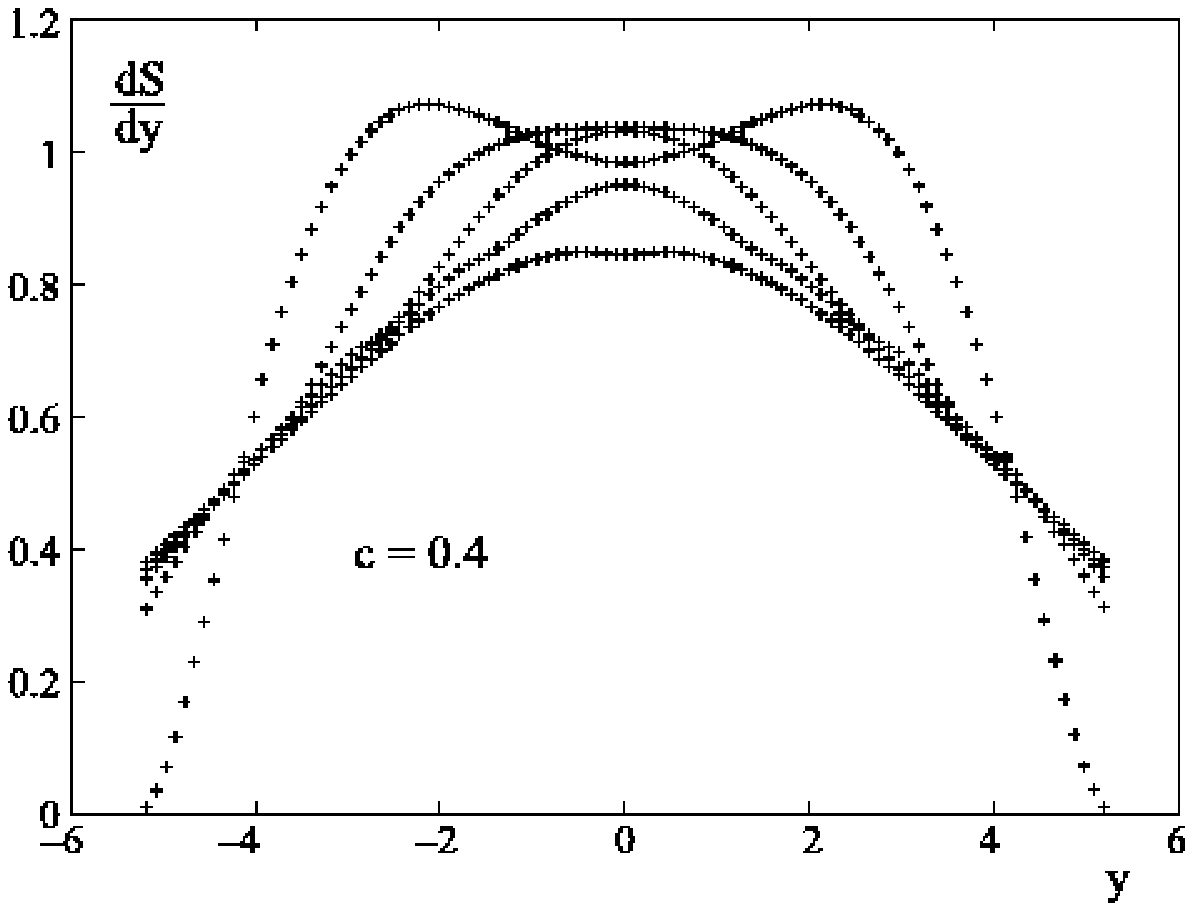}}
\caption{$dS/dy$ {\it for 5 growing values of 
$T_0/T_f=2^k, k=0 \to 4.$} $Left$: asymmetric one-component model;
 $ Right$: symmetric two-component model.
 Here one takes $c=0.4$ (see text). }
\lb{1}
\efig

In Fig. \ref{1}, we display the resulting curves corresponding to the
entropy $vs.$ rapidity distribution for 5 temperature values from the
initial one to the freeze-out, namely for $ T_0/T_f=2^k, k=0 \to 4.$ It can
be seen that the hydrodynamic diffusion towards the tails of the
one-component distribution (Fig. \ref{1}, left) is responsible for the
smooth disappearance of the two-bump initial structure of the
symmetrized distribution (Fig. \ref{1}, right) obtained by linear
combination. The initial bump structure may be easily modified by
changing the form of $h(x)$. This also modifies the evolution of the
entropy distribution as shown in Fig.\ref{2}. 

\bfig
{\includegraphics[height=0.25\textheight]{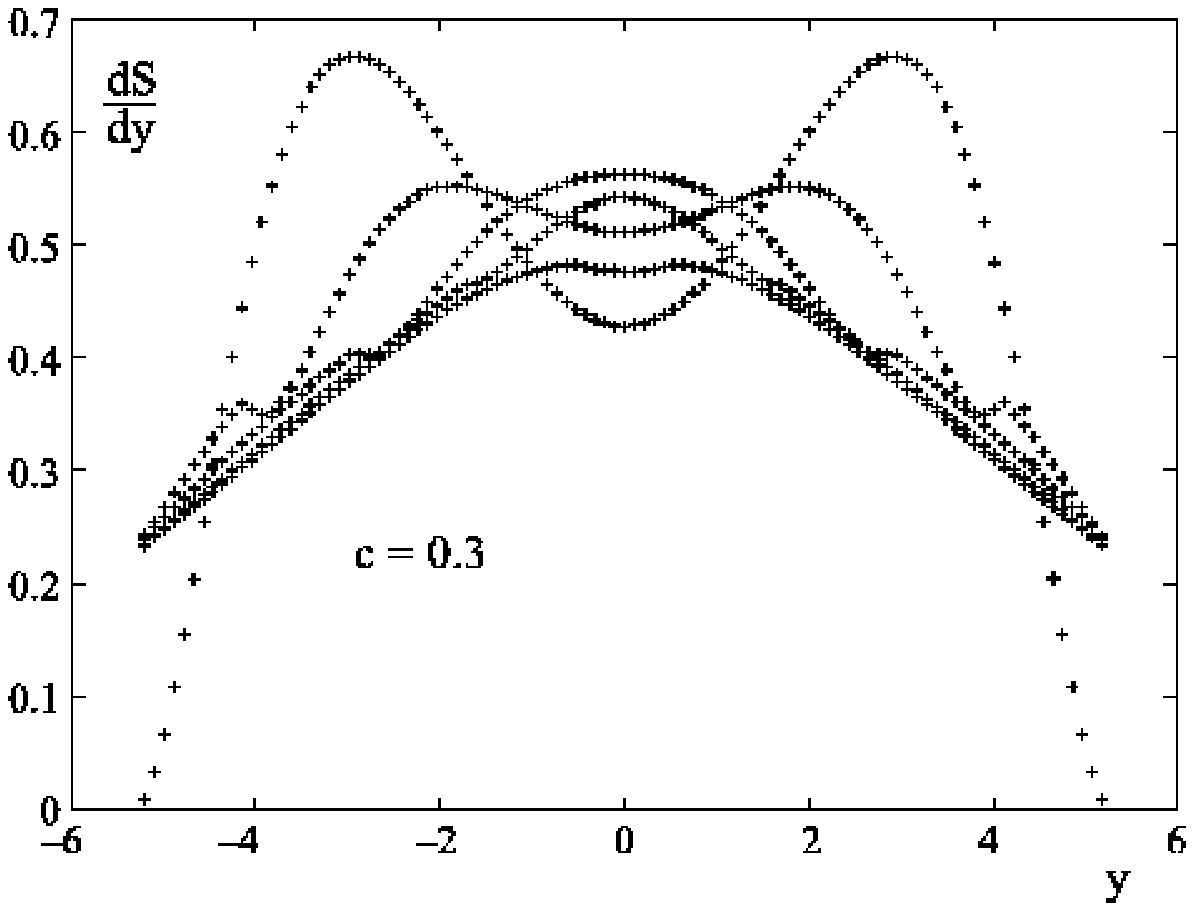}}
{\includegraphics[height=0.25\textheight]{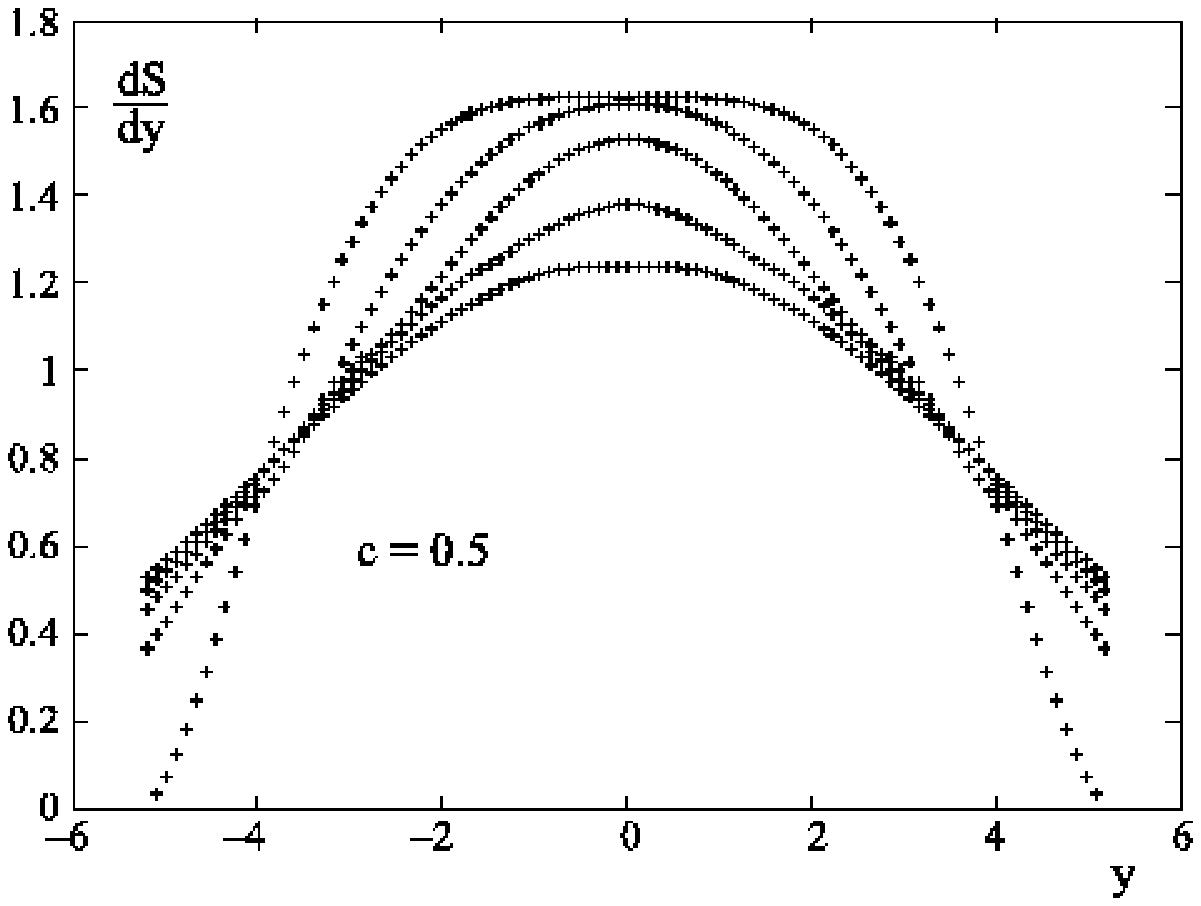}}
\caption{Two-component $dS/dy$ {\it for 5 growing values of 
$T_0/T_f=2^k, k=0 \to 4.$}
 $Left$:
 strongly bumped asymmetric input ($c=.3$);
 $Right:$ weakly bumped asymmetric input ($c=.5$). }
\lb{2}
\efig

It is also interesting to display some characteristic kinematic features
of the flow solutions, both for the asymmetric and symmetric cases. Note
that for those features, the symmetric configuration is $not$ a linear
symmetrized combination of both components, due to the nonlinear
character of the solution for kinematics \eqref{tt}. The
$\tau$-distribution  can be
found in Fig. \ref{3}.

\bfig
{\includegraphics[height=0.25\textheight]{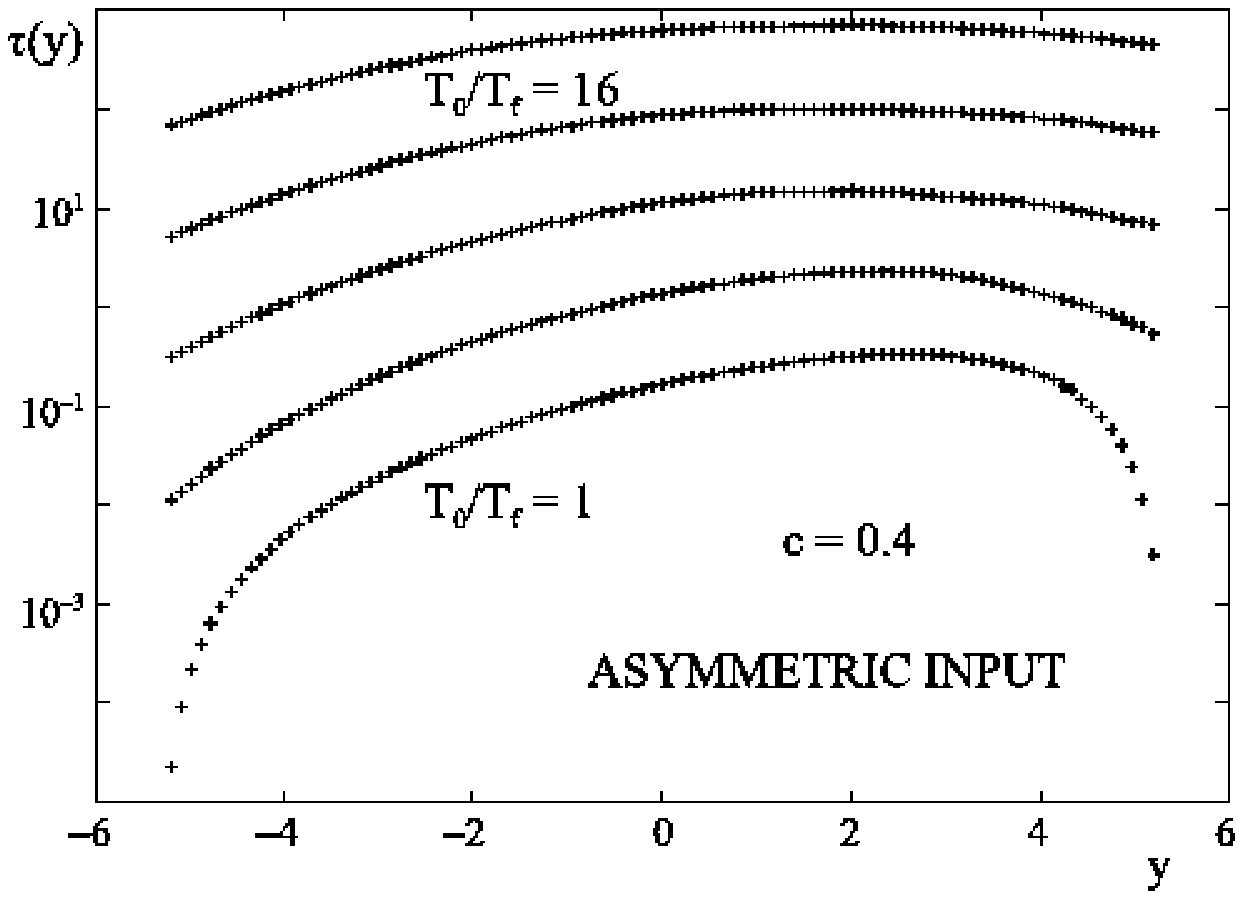}}
{\includegraphics[height=0.25\textheight]{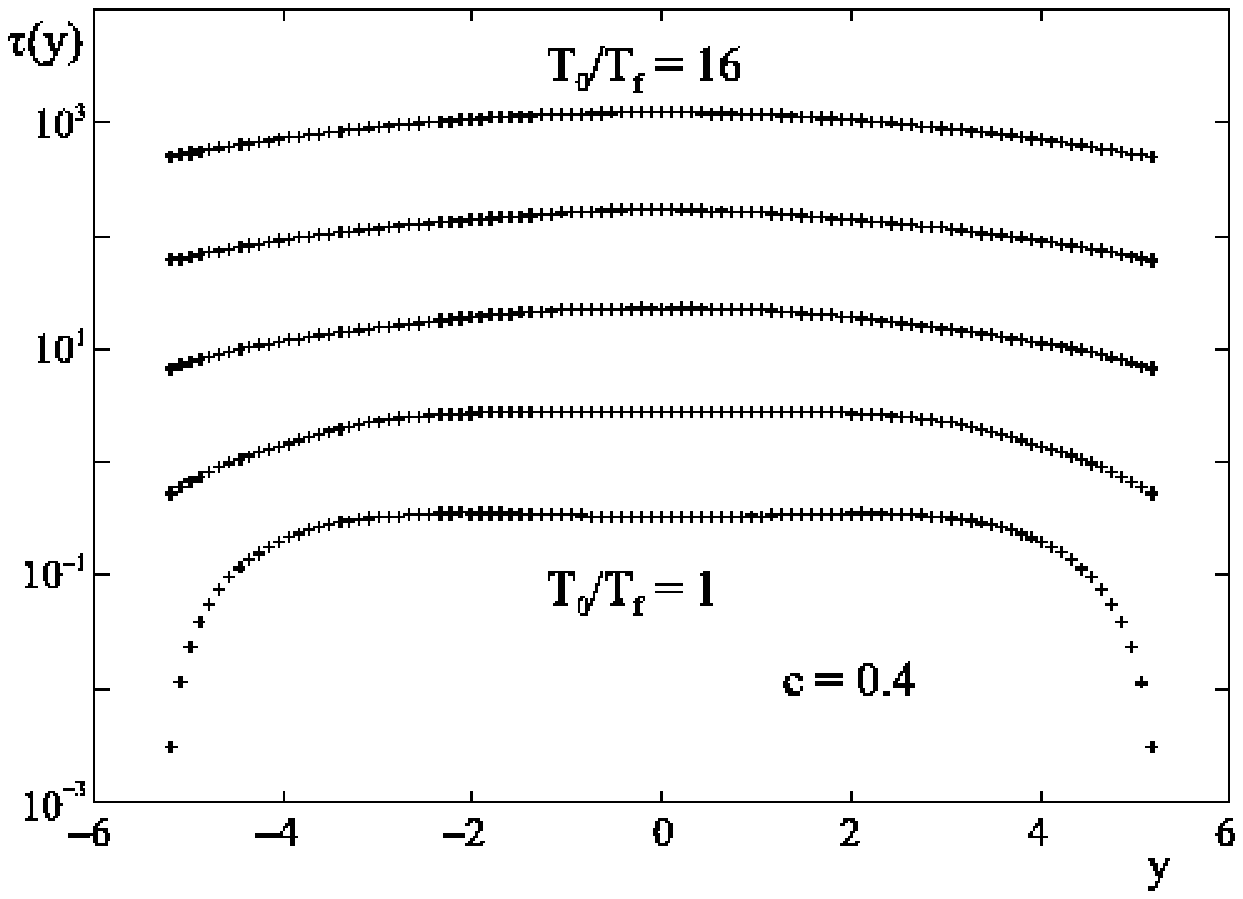}}
\caption{{\it Two-component proper-time distribution $\tau(x)$ 
 for 5 growing values of $T_0/T_f=2^k, k=0 \to 4.$} \newline  $Left$:
 asymmetric one-component model; $Right:$ symmetric two-component model.
 Here one takes $c=0.4$. }
\lb{3}
\efig

Concerning the kinematic properties of the
flow one can make the following comments.

First, as seen in the left panel of Fig.\ref{3}, the distribution of
proper times is significantly asymmetric. This means that, at any given
$y\neq 0$, the two asymmetric components freeze-out at different proper
times. This observation supports  the idea that the sets of particles
emitted by these two components are indepedent of each other (as required
by the analysis of the  long-range correlations \cite{bz,bz1,bzw}).
 
If the two components are mixed up during the evolution, the freeze-out
time is very different from the previous case. This is illustrated in
the right panel of Fig.\ref{3} where one sees that, for a fixed
temperature, the proper-time of the symmetric solution is close to
constant in the central rapidity region, at least at not-too-large
values of the ratio $T_0/T_f$. This is consistent with an approximate
Bjorken flow  \cite{bj}. One also observes that $\tau(y)$ increases with
increasing ratio $T_0/T_f$, as expected. Note that the curves for $T_0/T_f=1$ (the lowest ones in the Fig.\ref{3}) represent the proper time at which the evolution starts.

We have also evaluated how much the space-time rapidity $\eta$ deviates
from $y$ along the evolution. As shown in Fig. 4, we have found that the
difference $\eta-y$ is small, of the order of at most few \%. This
confirms that the the evolution of the system we are considering is
close to the Bjorken flow. All these features show that the flow is
rather smooth and laminar. We observe a slight but systematic tendency
to obtain $\eta \lesssim y,$ growing for large space-time rapidities,
which indicates that the flow is going ``outwards'', i.e. moving
apart on both sides from a kinematic tangential flow which would
correspond to the original ``In-Out'' cascade mechanism.

\bfig
{\includegraphics[height=0.25\textheight]{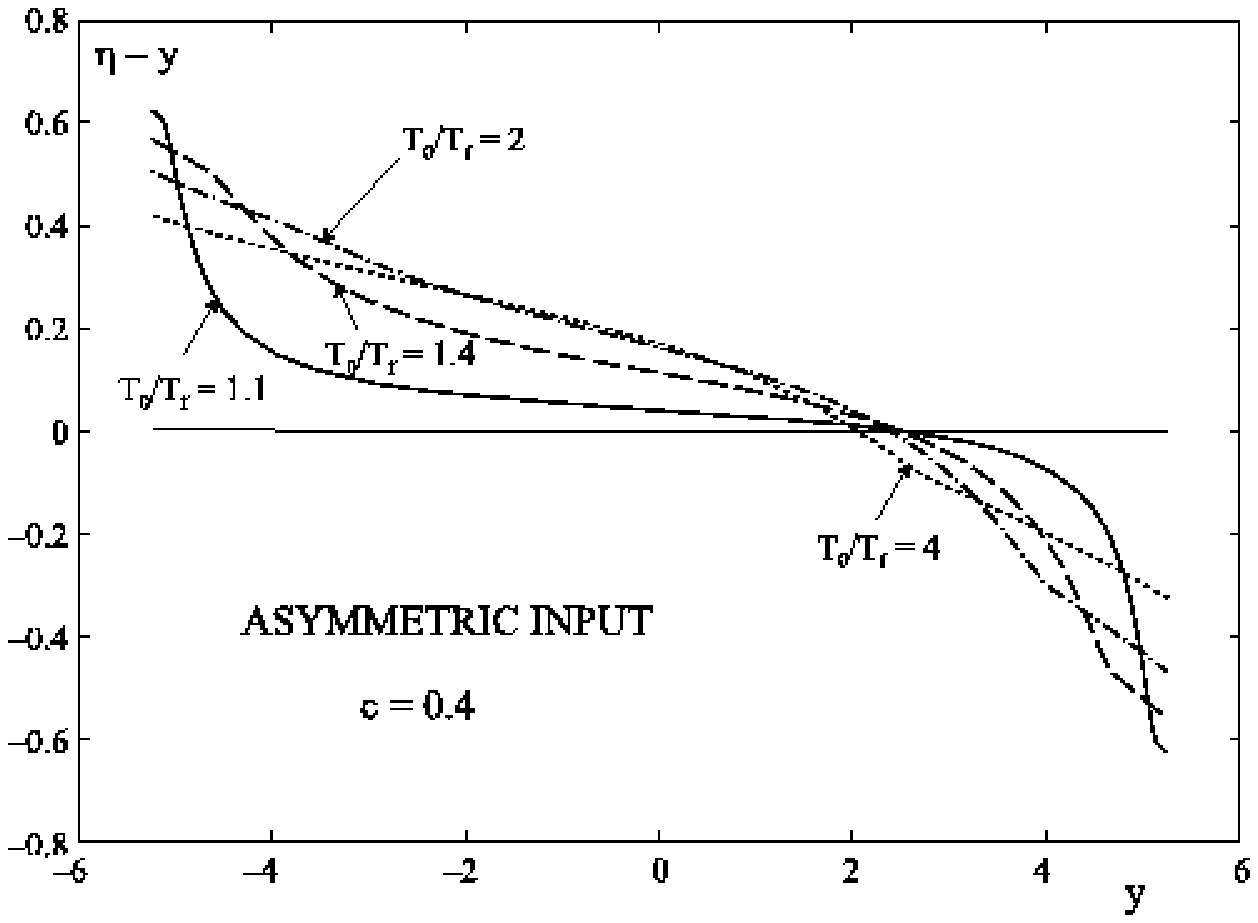}}
{\includegraphics[height=0.25\textheight]{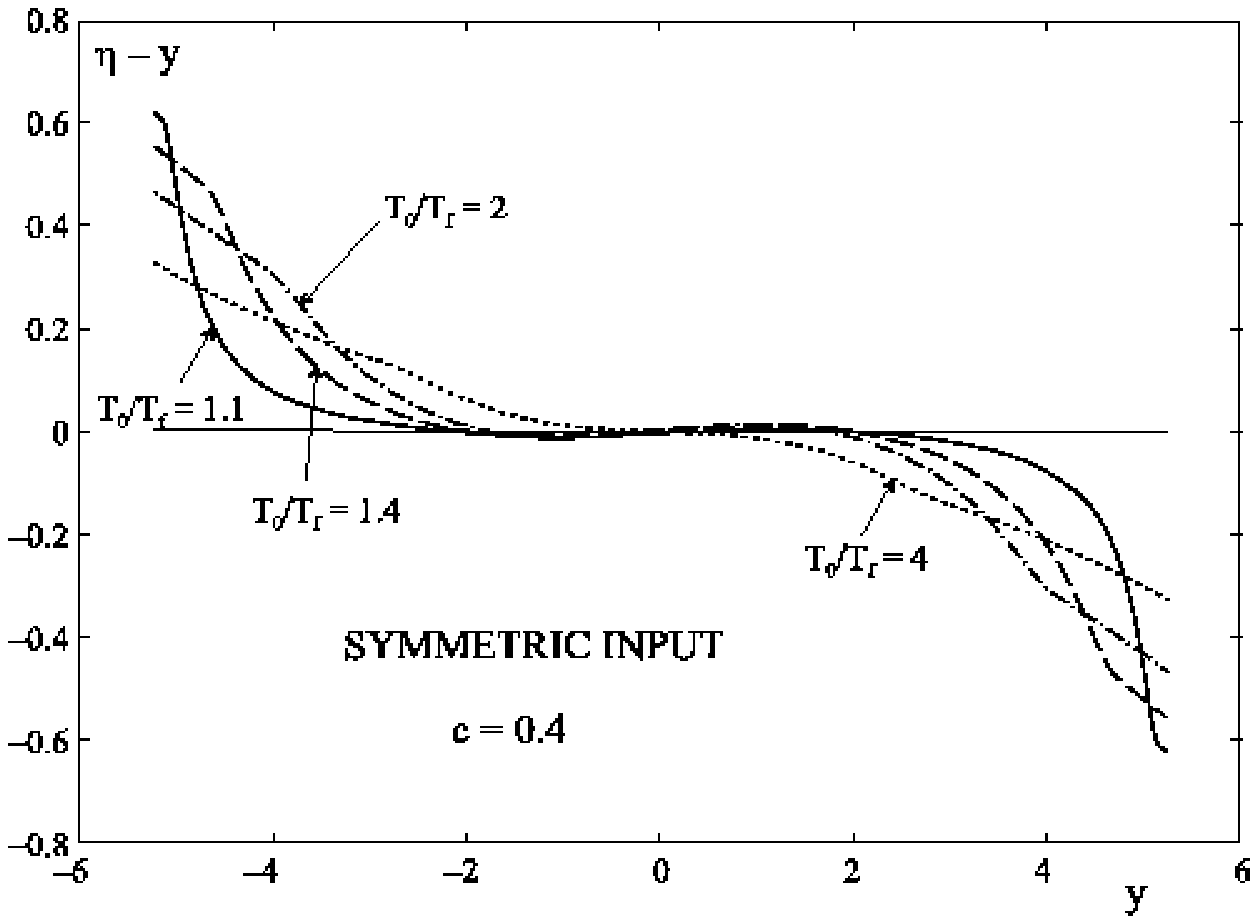}}
\caption{$\eta\!-\!y$ {\it for 5 growing values of 
$T_0/T_f$, marked in the figure; ($c=.4$). Note that $\eta-y=0$ for 
$T_0/T_f=1$}. 
$Left$: asymmetric one-component model;
 $ Right$: symmetric two-component model.}
\lb{4}
\efig

\section {Summary and comments}
\la{Summary}

In summary, we have investigated the longitudinal hydrodynamic
evolution of a one-dimetional perfect fluid (with a constant sound
velocity). Using the Khalatnikov equation in the light-cone variables,
derived earlier \cite{Beuf,ps}, the problem could be reduced to the
discussion of solutions of the (1+1) Klein-Gordon equation (called also
``Telegraphic equation'' \cite{bateman}). Using the well-known general form of these
solutions, we discussed their consequences in case the initial
conditions are asymmetric, as necessary e.g. for proton-nucleus or asymmetric nuclei
collisions and as also suggested by some models \cite{wbc, wbz,dpm,
fritjof}. 

We have studied in some detail only one simple but physically appealing example which can be
considered as a natural generalization of the Bjorken flow  \cite{bj} which, as is well-known, give a boost-invariant expansion of
the fluid. To relax boost-invariance, we assume that although at the
beginning of expansion $z^\pm \sim u^\pm$ (c.f. (\ref{zsimu})), the
proportionality coefficient is not a constant (as in the case of Bjorken
flow) but it is a function of $y$. This violates boost invariance 
and thus, among other rapidity configurations,  it is possible to study asymmetric ones. The results of our study may be synthesized as follows:

i) Our main conclusion is that the hydrodynamic expansion broadens rather
effectively the asymmetric bumpy structure of the entropy density (taken
for the input). Consequently, for symmetric collisions (as, e.g. $p\!-\!p$
 or $Pb\!-\!Pb$) where the asymmetric input has to be symmetrized, the
two-bump structure of the entropy density, characteristic of the
asymmetric input, changes easily into a single broad maximum. Thus we
have shown that even a strongly asymmetric input can be compatible with
data on particle production in symmetric processes. 

ii) We have also studied the properties of the flow. The most important
consequence of the initial conditions violating boost invariance is that
the proper time at which evolution begins is not constant but depends on
rapidity. In the simple case we considered, it is even downright
proportional to the initial entropy density. Also the freeze-out time
depends on rapidity, although the evolution has the tendency to smooth
out this dependence. 

iii) The rapidity dependence of $\tau_0$ and $\tau_f$ is especially
interesting for strongly asymmetric initial conditions. In this case
also the proper time is small in the region of small density and large
in the region of large density. Consequently, if two asymmetric sources
are present (as is the case for symmetric collisions) they freeze out at
different times (except at c.m. rapidity close to 0). Therefore one may
expect that they produce two bunches of particles which are indepenent of
each other.

iv) We have also studied the evolution of the difference $\eta\!-\!y$
(which vanishes at $t=0$ in our simple example). It turns out that this
difference remains small (below of 2-3 \% of $y$) throughout the
evolution (except at the very edge of phase-space where it may reach 10
\%). Thus we conclude that the flow has tendency to be laminar and
smooth, not very different from the Bjorken picture. It shows some
systematic tendency to behave ``outwards'' i.e. moving apart on both
sides from a kinematic tangential flow which would correspond to the
original ``In-Out'' cascade mechanism.

v) As an interesting by-product, we have found two general identities
valid for the (1+1)-dimensional approximation, corresponding to
conservation laws along the temperature (and thus proper-time) evolution
of the flow. One of them, see eq.\eqref{s}, can be interpreted as the
conservation of the full entropy. The other one, see eq.\eqref{ns}, up
to our knowledge is new, and still asks for a physical interpretation.

Let us add a comment on the microscopic interpretation of our results
for high-energy heavy ion collisions where the hydrodynamic set-up
proved strikingly pertinent for the description of the Quark-Gluon
Plasma phase of QCD \cite{Hydr,flor}. It is indeed extraordinary that
hydrodynamics describes successfully and almost unambiguously this
complicated process, belonging to the strongly coupled regime of QCD of
which one does not know yet a microscopic description (although the
string theory gives useful indications through the gauge/gravity duality
in a kinematic collisional framework \cite{string}, but for yet only
supersymmetric gauge field theories). 

As is well-known, the application of hydrodynamic description to heavy
ion collisions requires a rather rapid thermalization of the system of
quark and gluons created in the collision. This is not easy to reconcile
with the idea that this system is described by  { perturbative QCD} 
which is the usual scenario considered in this context\footnote{For other
approaches, see e.g. \cite{bfluc,bfch,fryb}.}. The perturbative QCD
evolution of the wave-function of a boosted projectile generates a set
of soft quantum fluctuations (gluons and $q\!-\!\bar q$ pairs) which
develop as a virtual cascading process. This is revealed, for instance,
when these {\it a priori} virtual fluctuations become observables, as in
the case of the ``hump-backed plateau'' \cite{dg} generated by a jet.
Similarly, such virtual fluctuations {play an important role in the theoretical understanding of deep-inelastic
reactions.} The problem arises because this system, { considered as the initial state of the incident nuclei} is highly anisotropic,
therefore its thermalization { after the collision} is difficult and { would} require relatively
long time.

Eventually, however, if the energy is sufficiently large, the soft
fluctuations are dense enough in phase-space to interact and eventually
lead to a saturation mechanism \cite{sat} grouping them around some
energy-dependent virtuality called the saturation scale. Therefore there
is a possibility that, due to soft interactions which are long range in
time, such a system thermalizes (at least approximately) already before
the { compound system created by the} collision takes place\footnote{It is worth to
mention that this scenario may also be valid in hadron-hadron and
hadron-nucleus collisions, when events of particularly high particle
density are considered.}. This is a possible model of the initial
conditions we considered in the paper. It naturally requires asymmetric
initial conditions. 

To conclude, our investigation shows that the longitudinal hydrodynamic
expansion of the system created in a high-energy collision may
significantly influence the original rapidity distribution of partons in
the colliding projectiles. It is therefore important to study this 
effect in more detail to obtain a reliable information on hadron
structure at high energies.

{ Acknowledgements}

This investigation was supported in part by the
grant N N202 125437 of the Polish Ministry of Science and Higher
Education (2009-2012). R.P. wants to thank the Institute of Physics for hospitality during the initial and final stages of the present work.

\end{document}